# Response to version 2 of the note concerning the observation of quantum Hawking radiation and its entanglement in an analogue black hole

Jeff Steinhauer

*Department of Physics, Technion—Israel Institute of Technology, Technion City, Haifa 32000, Israel*

The observation of quantum Hawking radiation and its entanglement in an analogue black hole was recently reported. A subsequent note (arXiv:1609.03803) criticized the study. Version 2 of the note was recently presented. The note suffers from technical difficulties which invalidate its main claims. We answer all of the comments in the note and show that the criticisms are not valid. We also answer a comment made by the author of the note in a different forum.

The main claims of the note [1] are based on analysis of the experimental results in the article [2]. The analysis in the note contains several errors. We first list the most severe errors and then explain how they result in the main conclusions of the note. Additional errors are discussed in the paragraphs after the list.

1. A set of experimentally measured values were literally replaced with zeros in the note. Specifically, the measured widths in $k$-space of the outgoing modes were replaced with zero for most points, as seen in the inset to Fig. 4a of the note. Furthermore, the actual non-zero values were used in a different part of the analysis, discussed in item 2.
2. Large error bars were added to the wavenumbers. In actuality, these discrete wavenumbers resulted from a Fourier transform of the experimental images. The wavenumbers are thus determined by the pixel size of the camera, the magnification of the imaging system, and the number of pixels used. The wavenumbers thus have no uncertainty other than the interval between adjacent wavenumbers. In other words, if a camera is used to image a noisy object, it does not cause the pixel size of the camera to become uncertain.
3. The note asserts that the population is measured to be strictly zero for large $k$, and not merely consistent with zero. In actuality, the measurement in no way shows a strictly zero population.

A main theme of the new version of the note is that the experiment reported in the article violates a theoretical bound, the Heisenberg limit. This bound implies that the correlations can be larger than the population by a large but limited factor. The note presents two arguments to show the violation of the bound. Firstly, the note asserts that the population is measured to be strictly zero for large $k$, and not merely consistent with zero. Thus, any correlations at large $k$ would violate the bound. In actuality, the measurement in no way shows a strictly zero population. Theoretically, even in the limit where dispersion can be neglected as in the gravitational case, one expects the population to be exceedingly small for a range of frequencies where the correlations are significant, but not strictly zero, as shown in Fig. 1c of the article.

Secondly, the note asserts that the finite width in $k$-space of the outgoing modes was not properly accounted for in computing the Heisenberg limit. However, the note uses the wrong experimental figure for the calculation (Fig. 3e rather than 3c of the article). More importantly, rather than using the widths appearing in Fig. 3e, the note replaces most of the widths with zero. This is seen in the inset to Fig. 4a of the note. Indeed, convoluting with a Gaussian of almost zero width will have little effect on the theoretical curve, resulting in a "violation" of the Heisenberg limit. In addition to these major difficulties, the note makes several less significant errors in computing the Heisenberg limit, which are discussed in the response to Comment 16 below.

In actuality, the Heisenberg limit is studied thoroughly in the article. The bound is determined from a preliminary experiment as well as the measurement of the population. It is seen in the article that the experiment is always within the bound (the solid curve is nowhere above the grey curve in Fig. 6a).

Another main assertion of the note is that the entanglement is statistically insignificant. However, this conclusion is based on a fundamental misunderstanding of the Fourier transform involved in the measurement. In a spatial Fourier transform, each $k$-bin has a certain width, and $k$ is known within that width. However, the note mistakenly multiplies this width by a factor of 2 to show "2 standard deviations", and mistakenly multiplies by another factor on the order of $\sqrt{2}$ since two $k$-variables appear in certain expressions. The width of each bin is mistakenly widened even further because the finite width in $k$-space of the modes is taken to be the width of the bin. Note that these widths of the modes are the very widths which were set to zero in the note, for the sake of computing the Heisenberg limit. Furthermore, the note uses the half width at half maximum as $1\sigma$, which further enlarges the bin by a factor of 1.2. In summary, the error bars on $k$ in the note are greatly exaggerated and meaningless. In contrast, the article shows the statistically robust entanglement with error bars in Fig. 6b. The caption points out that "the error bars shown are sufficiently spaced to be statistically independent."

Another point in the note is the suggestion that the Hawking radiation was measured at the borderline between weak and strong dispersion. In actuality, as stated in the article, "The horizon is hydrodynamic in the sense that its width is a few $\xi$, as shown in the inset of Fig. 2b." The inset also shows that the density distribution is similar to half a grey soliton, as predicted in [3]. The abstract of a recent work [4] studying the article states that such a profile, with parameters close to the experimental values, gives a spectrum which is "accurately Planckian in the relevant frequency domain, where the temperature differs from the relativistic prediction by less than 10%".

In another forum [5], the author of the note questioned whether the condensate used in the experiment is a true phase-coherent BEC, due to its 1D nature. The length scale of the phase coherence becomes short when the number of atoms per healing length is on the order of unity or less [6]. From the parameters given in the article, the number of atoms per healing length is 62 outside the sonic black hole and 27 inside. These values are sufficiently large to ensure phase coherence over lengths vastly longer than the condensate. Even if this were not the case, it has been suggested that Hawking radiation may still be observable [7].

**Response to all comments**

Each of the comments in version 2 of the note is given below, as well as a response. The list includes the main themes discussed above, as well as the other comments.

Comment 1: "The analysis raises severe doubts on the observation of Hawking radiation."

Response: This comment in the abstract is in no way supported by the rest of the note, as discussed below.

Comment 2: "The demonstration of black–hole lasing in Bose–Einstein condensates was recognised as a fluid–mechanical instability, although the author disputes that."

Response: This is not the case. Black-hole lasing (self-amplifying Hawking radiation) is indeed an instability [8]. In addition to this dynamical instability, the supersonic flow contains another instability (a zero-frequency mode). Both of these types of instability are clearly visible in the experiment and simulations [9].

Comment 3: "The experiment however, operates at the borderline between weak and strong dispersion"

Response: This is not the case. The inset to Fig. 2b of the article shows that the density distribution is similar to half a grey soliton, as predicted in [3]. The abstract of a recent work [4] studying the article states that such a profile, with parameters close to our experimental values, gives a spectrum which is "accurately Planckian in the relevant frequency domain, where the temperature differs from the relativistic prediction by less than 10%."

Comment 4: "the observed population distribution is clearly influenced by dispersion. Beyond a critical frequency, no radiation is measured."

Response: This comment does not reflect the experimental figures. Rather, it reflects a theoretical prediction. The experimental Fig. 5b in the article shows that the population becomes small within an error bar for large $k$.

Comment 5: "Yet the article [13] claims the observation of a Planck spectrum."

Response: This is not the case. As stated in the article, "The measured population is seen [in Fig. 5b] to agree well with the theoretical distribution of Hawking radiation". Furthermore, this theoretical distribution is described in the article, which states "Theoretical spectra … in which the Planck distribution is brought linearly to zero [3,10,11] at the measured $\omega_{peak}$". This is shown in the inset to Fig. 5b.

Comment 6: "Furthermore, the article reports on correlations of Hawking partners beyond the critical frequency, where there are no particles, which is impossible"

Response: This is not the case. The claim of the note that the population is strictly zero for large $k$ is completely unfounded (I assume that the comment refers to wavenumber, not frequency). The measurement in no way demonstrates a strictly zero population. Furthermore, the correlations beyond $k_{peak}$ are thoroughly explained in the article. Also, these correlations are observed in the numerical simulation, Fig. 6c. They are due to the finite $k$-width of the outgoing modes, which allows for $k$-components above $k_{peak}$. The components above $k_{peak}$ are verified in the preliminary oscillating horizon experiment (Figs. 3b, 3c, 3e), as well as the numerical simulation (Fig. 3f). The article states "[$k_{peak}$] has zero group velocity and diverging density of states per unit frequency, so the finite frequency width due to the finite duration of the experiment excites a broad wavepacket in $k$-space. Although the centre of the outgoing wavepacket must be below $k_{peak}$, components above $k_{peak}$ are expected. Indeed, the group velocity concept applies to wave packets rather than single modes. Such an outgoing wavepacket is clearly possible when viewed in the comoving frame in which $k_{peak}$ is not a special point."

Comment 7: "FIG. 1… error bars were taken from Fig. 3 of the article"

Response: This is not the case. "Figs. 3d and 3e of the article do not show error bars. Rather, they show the measured full width at half maximum of the $k$-distribution of the modes. As stated in the caption of Fig. 3, "The horizontal bars indicate the FWHM of the outgoing wavepacket."

Comment 8: "The article states: 'the Hawking distribution at low energies is thermal in the sense that the population goes like $1/\omega$'. ... There is no guarantee that the constant T in Eq. (5) has the meaning of a Hawking temperature (2) that is proportional to the velocity gradient. For example, an infinitely steep step from subsonic to supersonic speed would, according to Eq. (2), create an infinite Hawking temperature, yet due to dispersion the population is finite and behaves for small $\omega$ like Eq. (5) as well"

Response: This comment discusses issues not related to the sentence in the article, "the finite cross-section implies that the Hawking distribution at low energies is thermal in the sense that the population goes like $1/\omega$." The comment seems to be an attempt to extract the velocity gradient from the $1/\omega$ dependence. The comment points out that even an infinite velocity gradient would result in a finite Hawking temperature.

Comment 9: "In the experiment, the Hawking temperature is inferred from fitting the particle population obtained outside of the horizon with a Planck curve that is linearly brought to zero at $k_c$ (Fig. 2). The standard deviation of the fit, 0.025, lies within the error bar, 0.028, of the data, but a simple linear fit would give a standard deviation of 0.039, which is only marginally worse than the fit used to infer the Hawking temperature."

Response: In contrast to its own conclusion, this comment implies that a simple linear fit would give a poor fit. The fit would lie outside of the error bars. Looking at Fig. 5b of the article, it seems that a linear fit would indeed be poor.

Comment 10: "No measurement of the velocity gradient at the horizon was reported. Hence one cannot claim that the inferred T is indeed a Hawking temperature"

Response: The measurement was a direct measurement of the population, giving the Hawking temperature. Theoretically, the Hawking temperature should be approximately related to the gradients in the velocity and speed of sound.

Comment 11: "…nor that the spectrum is Planckian, as was claimed."

Response: This comment is not accurate. As stated in the article, "The measured population is seen [in Fig. 5b] to agree well with the theoretical distribution of Hawking radiation". Furthermore, this theoretical distribution is described in the article, which states "Theoretical spectra … in which the Planck distribution is brought linearly to zero [3,10,11] at the measured $\omega_{\text{peak}}$". This is shown in the inset to Fig. 5b.

Comment 12: "The figure shows no entanglement for $k_{\text{in}}\xi_{\text{in}} < 1.4$ inside the horizon that corresponds (Fig. 1) to $k_{\text{out}}\xi_{\text{out}} < 1.1$ outside the horizon where the agreement with the Planck curve (1) is best (Fig. 2)."

Response: Inspection of Fig. 6a of the article shows that this is not correct. Entanglement was seen for almost the entire $k$-range for which the population was measured (the dashed curve is almost always below the solid curve in Fig. 6a). The comment incorrectly implies that entanglement is only seen for the high-$k$ part of the dashed population curve.

Comment 13: "It does display entanglement for medium $k$, but then it goes on to show correlations beyond the critical $k_{\text{peak}}$ where no Hawking particles were observed, which is impossible."

Response A: This is not the case for two reasons. Firstly, the claim of the note that the population is strictly zero for large $k$ is completely unfounded. The measurement in no way demonstrates a strictly zero population. Secondly, the correlations beyond $k_{\text{peak}}$ are thoroughly explained in the article. Also, they are observed in the numerical simulation, Fig. 6c. These correlations are due to the finite $k$-width of the outgoing modes, which allows for $k$-components above $k_{\text{peak}}$. The components above $k_{\text{peak}}$ are verified in the preliminary oscillating horizon experiment (Figs. 3b, 3c, 3e), as well as the numerical simulation (Fig. 3f). The article states "[$k_{\text{peak}}$] has zero group velocity and diverging density of states per unit frequency, so the finite frequency width due to the finite duration of the experiment excites a broad wavepacket in $k$-space. Although the centre of the outgoing wavepacket must be below $k_{\text{peak}}$, components above $k_{\text{peak}}$ are expected. Indeed, the group velocity concept applies to wave packets rather than single modes. Such an outgoing wavepacket is clearly possible when viewed in the comoving frame in which $k_{\text{peak}}$ is not a special point."

Response B: Even in the limit where dispersion can be neglected in strict analogy with the gravitational case, one expects the population to be exceedingly small even for frequencies where the correlations are significant. This is seen in Fig. 1c of the article. The solid curve has fairly large values even where the dashed curve is very small. In other words, $|\alpha|^2|\beta|^2$ is much larger than $|\beta|^4$ for large frequencies, where $\alpha$ and $\beta$ are Bogoliubov coefficients.

Comment 14: "the correlation must vanish for vanishing $\bar{n}_{\text{H}}$ or $\bar{n}_{\text{P}}$: no correlation exists without population, yet the article shows correlations there (see the data reproduced in Fig. 3)."

Response: This comment does not reflect the data whatsoever. The population has error bars. There is no reason to suppose that the population is precisely zero above $k_{\text{peak}}$. The theoretical grey correlation curve of Fig. 6a of the article is derived from the measured Hawking temperature, obtained from the measured population. As expected, the measured correlations are never above this grey curve. Furthermore, even in the limit where dispersion can be neglected as in the gravitational case, one expects the population to be exceedingly small even for frequencies where the correlations are significant, but not strictly zero. This is discussed in Response B above.

Comment 15: "Relation (8) is called [13] the 'Heisenberg limit' (although it does not originate [19] from Heisenberg's uncertainty relation)."

Response: The Heisenberg limit is discussed as such in [12].

Comment 16: "Here the population curve $\bar{n}$ in $\bar{n}(\bar{n}+1)$ was convoluted with a Gaussian"

Response: This short comment is incorrect in three important ways. Firstly, the theoretical correlations (the Heisenberg limit) in the article were not convoluted with a Gaussian. Rather, they were convoluted with the measured $k$-distribution of the modes near $\omega_{\text{peak}}$ (Fig. 3c). Secondly, it was the theoretical correlation curve which was convoluted, not the population. The correlations are the directly observable quantity which is non-negligible at and above $k_{\text{peak}}$. Thirdly, the directly observable correlations are $S_0^2 |\langle \hat{b}_{k_{\text{HR}}} \hat{b}_{k_{\text{P}}} \rangle|^2$, where $S_0$ is the static structure factor which is proportional to $k$, and $\hat{b}_{k_{\text{HR}}}$ and $\hat{b}_{k_{\text{P}}}$ are annihilation operators for the Hawking and partner particles, respectively. Thus, the factor $S_0^2$ is included in the convolution. The unphysical calculation without $S_0^2$ performed in the note resulted in the divergence mentioned in Ref. 21 of the note.

Comment 17: FIG. 3 caption: "The populations vanish beyond $k_{\text{peak}}$ but not the correlations."

Response: This is not true. This is in no way seen in the data. The population has error bars. There is no reason to suppose that the population is precisely zero above $k_{\text{peak}}$. The theoretical grey correlation curve of Fig. 6a of the article is derived from the measured Hawking temperature, obtained from the measured population. As expected, the measured correlations are never above this grey curve. Furthermore, even in the limit where dispersion can be neglected as in the gravitational case, one expects the population to be exceedingly small even for frequencies where the correlations are significant, but not strictly zero. This is discussed in Response B above.

Comment 18: FIG. 3 caption: "The red curve shows the 'Heisenberg limit' $\bar{n}(\bar{n} + 1)$ with $\bar{n}$ obtained by convoluting the population curve with a Gaussian (Fig. 4)."

Response: The response to Comment 16 explains that this comment is incorrect in 3 important ways.

Comment 19: "…which increases the error bars of $k$ near $k_{\text{peak}}$ (Fig. 1)."

Response: Fig. 3d and 3e do not show error bars (uncertainty in $k$). Rather, they show the measured width of the $k$-distribution. There is a distribution of $k$ in the outgoing modes, including $k$-values above $k_{\text{peak}}$. It is not merely that we do not know the precise value of $k$ as implied in the comment.

Comment 20: "Convoluting the population curve $\bar{n}$ in $\bar{n}(\bar{n} + 1)$ with a Gaussian fits the last four data points of the correlation data if the standard deviation of the Gaussian is set to the constant 1.21."

Response: The response to Comment 16 explains that this comment is incorrect in 3 important ways.

Comment 21: "This $\sigma$ lies slightly above the maximal error bar 1.13 obtained from the dispersion measurements"

Response: This results from the incorrect calculation performed in the note. In the article on the other hand, no $\sigma$ is employed. Rather, the theoretical curve is convoluted with the measured $k$-distribution. Good agreement is seen between theory and experiment at high $k$ (the grey and solid curves in Fig. 6a of the article). Note that the correlations in Fig. 6a of the article end at $k_{\max}$, the maximum value seen in the preliminary experiment.

Comment 22: "The agreement with theory for the data points at the tail of the 'Heisenberg limit' is seen as the strongest evidence for the entanglement of Hawking radiation."

Response: This in no way reflects the text of the article. The article states that in Fig. 6b, "a substantial $k$-range of entanglement is seen". The discussion does not rely on the agreement with the Heisenberg limit. Entanglement requires that the correlation curve be above the population curve (the solid curve above the dashed curve in Fig. 6a). The last paragraph before the conclusion of the article discusses the statistical robustness of the entanglement. It is found that the probability of no entanglement is small. In summary, no entanglement would require much higher temperature than that observed, or a much wider correlation feature (much narrower in $k$-space).

Comment 23: "the Gaussian smoothing with maximal error bar is only justified in the vicinity of $k_{\text{peak}}$, as the dispersion measurements show."

Response: The note mistakenly introduces Gaussian smoothing. The article does not use Gaussian smoothing. Rather, the article convolutes the theoretical curve with the measured $k$-distribution of the outgoing modes.

Comment 24: "Taking the variation of $\sigma$ into account moves the curve of $\bar{n}(\bar{n}+1)$ below the correlation curve (Fig. 4): the observed correlations violate the 'Heisenberg limit' for large wavenumbers."

Response: The note replaces the experimental values of $\sigma$ with zero (!), resulting in a violation. When the convolution is performed with the actual data rather than zeros, the observation is within the Heisenberg limit (the solid curve is nowhere above the grey curve in Fig. 6a of the article). Furthermore, the convolution should be performed with the measured $k$-distribution, not with a Gaussian characterized by $\sigma$.

Comment 25: FIG. 4: "The fitted population curve is… convoluted with a Gaussian to produce the red curve…in agreement with the article."

Response: The article does not convolute with a Gaussian. Secondly, the correlations should be convoluted, since they are significant close to $k_{\text{peak}}$, as opposed to the population.

Comment 26: "The standard deviation of the Gaussian was set to the constant 1.21".

Response: Apparently, the value 1.21 is required to make the incorrect Gaussian convolution agree with the convolution with the measured $k$-distribution in the article.

Comment 27: The dotted curve shows the population curve convoluted with variable standard deviation $\sigma$ displayed in the insert"

Response: It is amazing that the $\sigma$ in the inset is zero for most of the frequency range! This in no way reflects the experimental values, as seen in Fig. 3e of the article. Indeed, convoluting with $\sigma = 0$ will have little effect on the curve.

Comment 28: "[the inset] reflects the actual uncertainty in $k_{\text{in}}\xi_{\text{in}}$ taken from the dispersion measurements"

Response: This is absolutely incorrect. The $\sigma = 0$ values in the inset are in sharp disagreement with the experimental values of Fig. 3e. The note sets the experimental values to zero. Furthermore, the experimental bars are not uncertainties. They are the full width at half maximum of the outgoing modes. The outgoing modes contain a broad distribution of $k$-values.

Comment 29: "Beyond the critical wavenumber $k_{\text{peak}}$ the correlation curve tends to lie above the corrected "Heisenberg limit" (dotted line), which violates the fundamental bounds"

Response: The note convolutes with zero rather than the experimental values, resulting in a "violation". A correct convolution, using the experimental $k$-distribution, shows no violation (the solid curve is nowhere above the grey curve in Fig. 6a of the article).

Comment 30: The two indications for Hawking radiation, the Planck spectrum for small wavenumbers and the alleged maximal entanglement for large wave numbers, are not consistent.

Response: This is incorrect in two senses. Firstly, these are not the two indications of Hawking radiation. Hawking radiation is observed by the correlations between the Hawking and partner particles, as seen in Fig. 4a of the article. Secondly, the comment seems to imply that the entanglement should be maximal everywhere, so the reduced entanglement observed for low wavenumbers is inconsistent with the population. In actuality, the reduced entanglement is interesting, but it is not inconsistent; there is no guarantee of maximal entanglement.

Comment 31: "Perhaps these inconsistencies disappear if all uncertainties in the experiment are taken into account"

Response: Firstly, there are no inconsistences, as discussed in the response to Comment 30. Furthermore, the note does not properly account for the experimental uncertainties. The measurement involves a spatial Fourier transform (the solid curve in Fig. 6a of the article is the Fourier transform of Fig. 4b). Each $k$-bin in a Fourier transform has a certain width. This width is the total uncertainty in $k$. It is determined by the pixel size of the camera, the magnification, and the number of pixels used. The note considers "$2\sigma$", and multiplies the width of each $k$-bin by a factor of 2. This step is not meaningful or justified mathematically. Furthermore, the note states that there are two $k$'s involved ($k_{\text{H}}$ and $k_{\text{P}}$), so the bins mistakenly get wider by another factor on the order of $\sqrt{2}$. In addition, the note mistakenly uses the observed width of the outgoing modes as the size of the bin, which further enlarges the bin. Note that these widths of the outgoing modes are the very widths which were set to zero in the note, for the sake of computing the Heisenberg limit. Furthermore, the note uses the half width at half maximum as $1\sigma$, which further enlarges the bin by a factor of 1.2. The resulting error bars, greatly enlarged relative to the actual bin size, are shown in the lower panel of Fig. 5 of the note. The article in contrast, shows the entanglement with error bars in Fig. 6b. The caption points out that "the error bars shown are sufficiently spaced to be statistically independent."

Comment 32: "FIG. 5: Entanglement with full error bars. The data of Fig. 3 is shown with full error regions (ellipses)."

Response: The horizontal error bars contain several mistakes. The error bars should be the bin size of the Fourier transform, but they have been greatly increased, as discussed in the response to the previous comment.

Comment 33: "The uncertainties in the variables $k_{in}\xi_{in}$ are obtained from Eq. (9) and the dispersion measurements"

Response: This comment implies three mistakes. Firstly, there is no reason to combine the uncertainties in $k$ as is done in Eq. 9. The correlations (the solid curve in Fig. 6a of the article) are obtained from the Fourier transform of Fig. 4b. The bin size of that Fourier transform is the uncertainty in $k$. The second mistake is that Fig 3d and 3e of the article do not show uncertainties in $k$. Rather, they show the measured full width at half maximum of the outgoing modes. Thus, using Fig. 3d and 3e exaggerates the uncertainty in k. Thirdly, the half width at half maximum of Figs. 3d and 3e were taken to be one standard deviation, which increases $\sigma$ by a factor of 1.2.

Comment 34: "the curves [of the lower panel of Fig. 5] are indistinguishable."

Response: The error bars computed in the note are quite meaningless, as explained in the response to Comment 31.

Comment 35: "The correlations $|\langle \hat{b}_H \hat{b}_P \rangle|^2$ and $\bar{n}_H \bar{n}_P$ are sensitive to statistical errors, as they depend on two wavenumbers, $k_H$ and $k_P$."

Response: This is not the case. Both $|\langle \hat{b}_H \hat{b}_P \rangle|^2$ and $\bar{n}_H \bar{n}_P$ are computed from Fourier transforms. The uncertainty in $k$ is the bin size of the Fourier transform.

Comment 36: "The error distribution of the wavenumber used for plotting is a convolution of the two individual error distributions. Assuming them to be Gaussian, the total variance is the sum of the two individual variances"

Response: This is not true for two reasons. Firstly, the uncertainty in $k$ is the bin size of the Fourier transform. It is not the sum of two contributions. Secondly, the "error distributions" used are not actually uncertainties. Rather, they are the measured $k$-distribution of the outgoing modes. These $k$-distributions were also measured with an uncertainty in $k$ given by the bin size.

Comment 37: "One reads off the error bars from the dispersion measurements and interpolates them."

Response: This is a mistake. As discussed in the previous comment, these "error bars" are actually the full width at half maximum of the $k$-distribution of the outgoing modes. They are not uncertainties. Furthermore, these are the very error bars which the note sets to zero in a different part of the analysis.

Comment 38: "Figure 5 shows the result: if the uncertainties of the wavenumbers are taken into account the data of the correlations is only distinguishable from the data of the populations squared for one standard deviation, the two data sets melt into each other for $2\sigma$."

Response: As discussed in the response to Comment 31, the horizontal error bars in Fig. 5 of the note are greatly exaggerated due to several mistakes. The error bars should be the bin size of the Fourier transform. Instead, the note computes them by combining two "error bars", giving a factor on the order of $\sqrt{2}$. Furthermore, the "error bars" used are not actually uncertainties. Rather, they are the widths of measured spectra. In addition, a factor of 1.2 is introduced due to confusing the half width at half maximum with $\sigma$. For the $2\sigma$ graph, an additional factor of 2 is mistakenly added.

Comment 39: "So, either one accepts inconsistencies in the data or the data become statistically insignificant."

Response: This is not the case. There are no inconsistencies, as discussed in the response to Comment 30. The assertion of the note that the data is statistically insignificant is based on a mathematically invalid and meaningless analysis, as explained in the response to Comment 31. In contrast, the article explains that the observation of entanglement is statistically robust. The article explains, "The probability of no entanglement is small. First, the $0.8\ mc_{\text{out}}^2$ dotted curve in Fig. 6a indicates a temperature that would have resulted in a substantially reduced entanglement region, if it had been observed. However, such a high temperature is seen to be ruled out by the measurement in Fig. 5a. Second, if $S_0^2 |\langle \hat{b}_{k_{HR}} \hat{b}_{k_P} \rangle|^2$ in Fig. 6a were narrower by a factor of 0.53, then there would have been no entanglement. This would require that the profile of Fig. 4b would be wider by 90%, but the uncertainty in the width is only 16%, as determined by a least-squares fit of a Gaussian including the contribution of the experimental error bars. In contrast to the entanglement seen for the Hawking radiation, the oscillating horizon experiment shows classical correlations ($\Delta > 0$), as expected (Fig. 3h)." Furthermore, Fig. 6b of the article shows a substantial $k$-range of entanglement, and its caption explains that "the error bars shown are sufficiently spaced to be statistically independent."

Comment 40: "Additionally, assumptions were made throughout the data analysis based on the expectation of Hawking radiation."

Response: This is not the case. The assumptions made are clearly stated in the article, even in the abstract.

Comment 41: "For quantifying the degree of entanglement, the correlation was compared with the population of only one of the Hawking partners (the one outside the horizon) and not also with the population of the other... For analysing the entanglement, it was assumed that $\bar{n}_\text{H}$ and $\bar{n}_\text{P}$ were the same. However, this was part of Hawking's prediction, and cannot be taken for granted."

Response: The measured population is so small that even a significant difference between the Hawking and partner populations would not be sufficient to imply no entanglement. This is discussed in the paragraph before the conclusion in the article, which states that the very high temperature of $0.8\ mc_\text{out}^2$ would be required to substantially reduce the entanglement.

Comment 42: "Assumptions were also made in the method of obtaining the particle correlations from the experimental data. The correlations were inferred from subsets of the data integrated along lines parallel to a line of expected correlations, a line found by optimisation. This method selects the Fourier–components of the density–density correlations evaluated within a subset of the data that would be consistent with the expected particle correlations. All other Fourier–components were ignored. Although this would give the Hawking correlations if they are there, it does not allow a comparison with the level of the other Fourier components: it does not discriminate between signal and noise, nor between signal and background."

Response: This comment does not reflect the data analysis employed. The first step of the analysis was to objectively find the angle of the correlation feature by the inset of Fig. 4b in the article. The profile at that angle and along the entire correlation feature was then computed, as shown in Fig. 4b. It was found that the angle was in agreement with the preliminary oscillating horizon experiment, as shown in Fig. 3g. Eq. 2 is then used to compute the correlations.

Comment 43: "The method of extracting correlations [23] involves another vital assumption: it assumes that only the Hawking partners are generated in accelerating the condensate beyond the speed of sound, but no other excitations in the fluid."

Response: This is not correct in two senses. Most importantly, the discussion in the article includes additional insights relative to Ref. 23 of the note. The article makes no assumption about which modes are populated. Rather, it is assumed that modes of different frequencies are not correlated. As stated in the abstract, the entanglement is measured "within the reasonable assumption that excitations with different frequencies are not correlated". The assumption is explained in the article: "The frequencies of [the negative-$k$] phonons are increased by $2|k|v$ relative to the positive-$k$ phonons, due to the Doppler shift. Thus, the neglected terms represent correlations between widely separated phonons on opposite sides of the horizon with different frequencies. Hawking radiation would not create such correlations, and it is not clear what would. Thus, (1) should be a good approximation in the general case of spatially separated regions in an inhomogeneous, flowing condensate. The reasonable assumption that the terms can be neglected allows for our measurement of the entanglement." Furthermore, the lack of correlations between phonons of different frequencies is proven in [10] for the stationary case. The other sense in which the comment is not quite correct is that Ref. 23 only neglects the negative-$k$ phonons (traveling with the flow).

Comment 44: "As justification, the article states: 'the neglected terms represent correlations between widely separated phonons on opposite sides of the horizon with different frequencies.' However, this statement contradicts the dispersion measurements. Figure 1a shows the wavenumbers and frequencies for the two possible modes of excitations inside the horizon (represented by filled and open circles in the figure): the Hawking waves attempting to travel against the flow and waves traveling with the flow. One sees that their wavenumbers and frequencies are similar."

Response: This is not the case. The comment's mention of "similar" frequencies is a qualitative and irrelevant statement. The frequencies are in no way the same. Specifically, the correlations are measured for a given value of $k$. As stated in the article, "the frequencies of [the negative-$k$] phonons are increased by $2|k|v$ relative to the positive-$k$ phonons, due to the Doppler shift." This Doppler shift is clearly seen in the dispersion relation (Fig. 3e of the article). For a given $k$-value, there is a vertical shift between the dashed and solid curves.

Comment 45: "Neglecting the influence of extra excitations on the density–density correlation amounts to an assumption that has not been independently verified in the article."

Response: The assumptions made are clearly stated in the article. As stated in the abstract, the entanglement is measured "within the reasonable assumption that excitations with different frequencies are not correlated". The assumption is explained in the article: "The frequencies of [the negative-$k$] phonons are increased by $2|k|v$ relative to the positive-$k$ phonons, due to the Doppler shift. Thus, the neglected terms represent correlations between widely separated phonons on opposite sides of the horizon with different frequencies. Hawking radiation would not create such correlations, and it is not clear what would. Thus, (1) should be a good approximation in the general case of spatially separated regions in an inhomogeneous, flowing condensate. The reasonable assumption that the terms can be neglected allows for our measurement of the entanglement." Furthermore, the lack of correlations between phonons of different frequencies is proven in [10] for the stationary case.

Comment 46: "This paper has analysed the evidence for the observation of quantum Hawking radiation and its entanglement in the recent article and found severe problems on several accounts."

Response: The criticisms in the note are refuted above.

Comment 47: "First, the observed spectrum of Hawking radiation is not Planckian, although the paper claims it to be."

Response: This is not the case. As stated in the article, "The measured population is seen [in Fig. 5b] to agree well with the theoretical distribution of Hawking radiation". Furthermore, this theoretical distribution is described in the article, which states "Theoretical spectra … in which the Planck distribution is brought linearly to zero [3,10,11] at the measured $\omega_{\text{peak}}$". This is shown in the inset to Fig. 5b.

Comment 48: "Second, particle correlations are observed without particles being present,"

Response: This comment does not reflect the data whatsoever. The population has error bars. There is no reason to suppose that the population is strictly zero for large $k$. The theoretical grey correlation curve of Fig. 6a of the article is derived from the measured Hawking temperature, obtained from the measured population. As expected, the measured correlations are never above this grey curve. Furthermore, even in the limit where dispersion can be neglected as in the gravitational case, one expects the population to be exceedingly small even for frequencies where the correlations are significant, but not strictly zero. This is discussed in Response B above.

Comment 49: "unless, third, wavenumber uncertainties reduce the statistical confidence to one standard deviation, i.e. to insignificance."

Response: This is not true. The "wavenumber uncertainties" found in the note reflect a fundamental misunderstanding of the Fourier transform involved in the measurement. In a spatial Fourier transform, each $k$-bin has a certain width, and $k$ is known within that width. However, the note mistakenly multiplies this width by a factor of 2 to show "2 standard deviations", and mistakenly multiplies by another factor on the order of $\sqrt{2}$ since two $k$-variables appear in certain expressions. The width of each bin is mistakenly widened even further because the finite width in $k$-space of the outgoing modes is taken to be the width of the bin. Furthermore, the note uses the half width at half maximum as $1\sigma$, which further enlarges the bin by a factor of 1.2. Note that these widths of the outgoing modes are the very widths which were set to zero in the note, for the sake of computing the Heisenberg limit. In summary, the error bars on $k$ in the note are exaggerated and meaningless. In contrast, the article shows the statistically robust entanglement with error bars in Fig. 6b. The caption points out that "the error bars shown are sufficiently spaced to be statistically independent."

Comment 50: "Fourth, no measurement of the population of Hawking partners was reported."

Response: The population of Hawking particles is measured in the article, and is seen in Fig. 5b. As stated in the article, the populations of Hawking and partner particles should be the same.

Comment 51: "Fifth, the method of inferring particle correlations from the measured density–density correlations relies on the expectation that Hawking radiation is present and nothing else."

Response: This is not the case. The article makes no assumption about which modes are populated. Rather, it is assumed that modes of different frequencies are not correlated. As stated in the abstract, the entanglement is measured "within the reasonable assumption that excitations with different frequencies are not correlated". The assumption is explained in the article: "The frequencies of [the negative-$k$] phonons are increased by $2|k|v$ relative to the positive-$k$ phonons, due to the Doppler shift. Thus, the neglected terms represent correlations between widely separated phonons on opposite sides of the horizon with different frequencies. Hawking radiation would not create such correlations, and it is not clear what would. Thus, (1) should be a good approximation in the general case of spatially separated regions in an inhomogeneous, flowing condensate. The reasonable assumption that the terms can be neglected allows for our measurement of the entanglement." Furthermore, the lack of correlations between phonons of different frequencies is proven in [10] for the stationary case.

Comment 52: "Overall, an unbiased, blind analysis of the data was not reported, which does not conform to the standards of claiming a discovery."

Response: I believe that the article employed techniques giving a level of confidence beyond what is typical for the field of ultracold atoms. Firstly, a preliminary experiment was created and performed to verify the nature of the outgoing modes which one would expect in the Hawking radiation. Secondly, a numerical simulation including quantum fluctuations verified the results. Thirdly, there is a paragraph on the last page of the article analyzing the statistical confidence of the observation of entanglement. It is found that the entanglement is confirmed with a significance of 90/16 = 5.7 standard deviations. Fourthly, the work reflects the input and questions of many colleagues. After the original version of the article appeared on the arXiv, there was a meeting in Paris to discuss the article. There I received the input of several colleagues, and their questions were studied and included in the later versions. Furthermore, the 3 referees provided invaluable comments reflected in the later versions.

Comment 53: "Instead of attempting to fully confirm Hawking's theory in the laboratory, the experiment may have missed discovering a different interesting phenomenon, like its predecessor."

Response: This is not the case. Black-hole lasing (self-amplifying Hawking radiation) was observed in [13]. In addition to this dynamical instability, the supersonic flow contains an energetic instability (a zero-frequency mode). Both of these types of instability are clearly visible in the experiment and simulations [9].

Comment 54: "seeding the Hawking process with a coherent state does produce entanglement. Such situations may mimic Hawking radiation in some aspects, but not in others, which, apart from statistical uncertainty, might explain the puzzling results of the article."

Response: This is not in agreement with the current understanding. The abstract of [14] states that "nonseparability of the phonon modes offers an unambiguous signature of the quantum origin of the phonon emission".

Comment 55: "While the experimental data of the article are valid, although incomplete, the conclusions and claims are not"

Response: The experimental data is complete and the conclusions and claims are valid, as discussed in the responses to the comments above.

Comment 56: "…especially the claim of having observed entanglement with a statistical confidence of 90/16 = 5.7$\sigma$, which reduces to the order of 1$\sigma$ on closer inspection."

Response: This comment is based on the note's mathematically invalid and meaningless analysis, as explained in the response to Comment 31. In contrast, the article explains that the observation of entanglement is statistically robust. The article explains, "The probability of no entanglement is small. First, the 0.8 $mc_{\text{out}}^2$ dotted curve in Fig. 6a indicates a temperature that would have resulted in a substantially reduced entanglement region, if it had been observed. However, such a high temperature is seen to be ruled out by the measurement in Fig. 5a. Second, if $S_0^2 |\langle \hat{b}_{k_{HR}} \hat{b}_{k_P} \rangle|^2$ in Fig. 6a were narrower by a factor of 0.53, then there would have been no entanglement. This would require that the profile of Fig. 4b would be wider by 90%, but the uncertainty in the width is only 16%, as determined by a least-squares fit of a Gaussian including the contribution of the experimental error bars. In contrast to the entanglement seen for the Hawking radiation, the oscillating horizon experiment shows classical correlations ($\Delta > 0$), as expected (Fig. 3h)." Furthermore, Fig. 6b of the article shows a substantial $k$-range of entanglement, and its caption explains that the error bars shown are sufficiently spaced to be statistically independent.

Comment 57: "One cannot claim that quantum Hawking radiation and its entanglement has been observed by the standards of a discovery. What has been observed is a matter of debate."

Response: I believe that the article employed techniques giving a level of confidence beyond what is typical for the field of ultracold atoms. Firstly, a preliminary experiment was created and performed to verify the nature of the outgoing modes which one would expect in the Hawking radiation. Secondly, a numerical simulation including quantum fluctuations verified the results. Thirdly, there is a paragraph on the last page of the article analyzing the statistical confidence of the observation of entanglement. It is found that the entanglement is confirmed with a significance of 90/16 = 5.7 standard deviations. Fourthly, the work reflects the input and questions of many colleagues. After the original version of the article appeared on the arXiv, there was a meeting in Paris to discuss the article. There I received the input of several colleagues, and their questions were studied and included in the later versions. Furthermore, the 3 referees provided invaluable comments reflected in the later versions.

I thank Renaud Parentani, Amos Ori, and Joseph Avron for helpful comments. This work was supported by the Israel Science Foundation.